\documentstyle[epsfig]{mn2e}

\newif\ifAMStwofonts

\title[Limit cycle instability in magnetized accretion discs]
      {On the limit cycle instability in magnetized accretion discs}

\author[Merloni \& Nayakshin]{Andrea Merloni$^{1}$ \& Sergei Nayakshin$^{2}$\\$^{1}$Max-Planck-Institut f\"ur Astrophysik,
Karl-Schwarzschild-Strasse 1, D-85741, Garching,
Germany\\$^{2}$Department of Physics and Astronomy, University of
Leicester, Leicester, LE1 7RH, UK}

\date{}

\def\ledd{L_{\rm Edd}}

\newcommand\fr{F_{\rm rad}}

\newcommand\pr{P_{\rm rad}}

\newcommand\pg{P_{\rm gas}}

\newcommand\ptot{P_{\rm tot}}

\newcommand\simlt{\lower.5ex\hbox{$\; \buildrel < \over \sim \;$}}

\newcommand\simgt{\lower.5ex\hbox{$\; \buildrel > \over \sim \;$}}

\begin{document}

\maketitle

\label{firstpage}

\begin{abstract}
Observational evidence accumulated over the past decade indicates that
accretion discs in X-ray binaries are viscously stable 
unless they accrete very close to the Eddington limit. 
This is at odds with the most basic standard accretion disc theory, but
could be explained by either having the discs to be much cooler whereby they
 are not radiation pressure dominated, or by a more sophisticated viscosity
law. Here we argue that the latter is taking place in practice, on the
basis of a stability analysis that assumes that the
magneto-rotational-instability (MRI) responsible for generating the
turbulent stresses inside the discs is also the source for a
magnetically dominated corona. We show that observations of
stable discs in the high/soft states of black hole binaries, on the one hand,
and of the strongly variable microquasar GRS 1915+105 on the other, can all be
explained if the magnetic turbulent stresses inside the disc scale
proportionally to the geometric mean of gas and total pressure with a
constant of proportionality (viscosity parameter) having a value of a
few times $10^{-2}$. Implications for bright AGN are also briefly discussed.
\end{abstract}


\section{Introduction}
According to the widely adopted standard solutions (Shakura \& Sunyaev 1973),
 luminous accretion discs close to the Eddington rate should be radiation
 pressure dominated and therefore unstable to perturbations of both mass flow
 (Lightman \& Eardley 1974) and heating rate (Pringle, Rees \& Pacholczyk
 1973) for the commonly adopted assumption that viscous stresses within the
 disc are proportional to the total (gas$+$radiation) pressure
 ($\alpha$-viscosity prescription).  Taking into account the stabilizing
 effect of radial advection near the Eddington rate,
 often modeled with the ``slim disc'' solutions (Abramowicz et
 al. 1988), a limit cycle-type of behaviour should be expected, which has been
 confirmed by numerical simulations of time-dependent discs (Taam \& Lin 1984;
 Honma, Matsumoto \& Kato 1991; Szuszkiewicz \& Miller 1997).

These instabilities may also operate in accretion discs of supermassive black
holes. It is now believed that the main phase of growth of these black holes
must have occurred in short-lived episodes of near-Eddington accretion (Yu \&
Tremaine 2002; Merloni 2004; Hopkins et al. 2006), most likely associated with
bright quasar phases.  The lifetimes and duty-cycles
of such luminous objects, crucial to understand their cosmological evolution,
may be influenced by the radiation-pressure instabilities (or lack thereof) of
accretion discs.

Galactic black holes in binary systems provide an important laboratory to study
these instabilities on humanly observable time scales.  Among all of them, only
one -- the famous microquasar GRS 1915+105 -- displays a variability which can
be explained by some kind of viscous instability (Belloni et al. 1997;
Nayakshin, Rappaport \& Melia 2000; Janiuk, Czerny \& Siemiginowska 2000; for
a recent review and a more comprehensive list of references on this enigmatic
source, see Fender \& Belloni 2004).  On the other hand, the vast majority of
transient galactic black holes are observed to be stable in a disc-dominated
state (the so-called high/soft state, or thermal dominant state, see
McClintock \& Remillard 2006) at luminosities which are a few tenths of the
Eddington one, which is hard to reconcile with the original viscosity
prescription of stresses proportional to total pressure.

Two physically plausible ways to make luminous accretion disks stable are well
known. First of all, the discs may be cooling much faster than the standard
solution assumes due to an additional rapid energy transfer from the disc
mid-plane into a corona, a jet or a wind, so that
the radiation pressure simply never dominates in the disk (e.g., Svensson \&
Zdziarski 1994). Secondly, the
anomalous viscosity of accretion discs, now understood to be due to MRI, may
not scale with the total disc pressure (e.g., Lightman \& Eardley 1974).  Numerical MHD simulations of
turbulent accretion flows are the most promising tools for differentiating
between these possibilities from first principles (see, in particular, Sano
et al. 2004 and references therein). However, due to immense numerical
challenges, one will have to wait until global 3-D radiative simulations are
performed over a large enough range of parameter space before a clear answer
will emerge.

In this paper we investigate the
time-dependent evolution of magnetized accretion discs by means of
semi-analytic techniques and discuss their relation to observations. In
particular, we suggest plausible 
scalings for the fraction of power
dissipated in the corona and the viscosity law, and we perform a
local viscous disc stability analysis, later checked with
time-dependent disc models. We find that the observed stability properties of
galactic black hole discs are most likely explained by a modified viscosity
law rather than by an extra coronal or jet cooling.

\section{Viscosity law}

\label{sec:visc}

The nature and extent of the posited limit cycle instabilities at high
accretion rates depend critically on the
poorly understood prescription for the viscous torques: it is well
 known that assuming viscous stresses scale proportionally to the
 gas pressure results in accretion discs which are stable
 throughout, even at the highest accretion rates (Lightman and Eardley
 1974; Stella and Rosner, 1984).
In fact, our ignorance of the physical mechanisms giving rise to
the disc viscosity, and in particular of its exact scaling, has led 
many authors to consider the  outcome of radiation
pressure dominated discs obeying a more general prescription for the
viscous stresses $t_{r\phi}$ \cite{tl84,szu90,hmk91,wm03}:

\begin{equation}
\label{eq:defmu}
t_{r\phi}=\alpha_0 P_{\rm tot}^{1-\mu/2} P_{\rm gas}^{\mu/2},
\end{equation}
with $\alpha_0$ constant, where 
\begin{equation}
\label{eq:state}
P_{\rm tot}=P_{\rm gas}+P_{\rm rad} = 2 \rho\, m_p^{-1} kT + a T^4/3
\end{equation} 
is the sum of gas plus radiation pressure (for hydrogen rich material), 
 $T$ is the mid-plane disc temperature and $m_p$ is the proton mass. 
Within this approach, the parameter $\mu$ can take any value between 0
(stresses proportional to total pressure) and 2 (stresses
proportional to gas pressure only).

\subsection{Magneto-Rotationally Unstable discs and the generation of
 the corona}
Numerical studies of the last decade have shed new light on the nature of 
viscosity in accretion discs, by elucidating the crucial role of MHD
turbulence for their enhanced transport properties. Since magneto-rotational
instability (MRI; see Balbus \& Hawley 1998, and reference therein) is the
primary driver of the angular momentum transfer in the discs, the turbulent
magnetic stresses scale with magnetic pressure, and therefore $t_{r \phi}
\propto P_{\rm mag}$, where $P_{\rm mag} = |B_{\rm disc}|^2/8\pi$ is the
magnetic pressure inside the disc.

One of the main open issues in the physics of black hole accretion discs is
the relationship between the disc MRI-driven turbulent viscosity and the
generation of the hot coronae that are usually postulated in order to explain
the observed X-ray emission (Liang \& Price 1977; Galeev, Rosner \& Vaiana
1979; Blackman \& Field 2000; Kuncic \& Bicknell 2004).  Phenomenological
models usually assume that at each radius, a fraction $f$ of the internally
generated power is transferred vertically outside the disc, 
and powers a magnetically dominated corona (Haardt \& Maraschi
1991; Svensson \& Zdziarski 1994). As customary (see e.g. Svensson 
\& Zdziarski 1994; Merloni 2003), we assume that in MRI-turbulent
discs such a fraction $f$ of the binding energy is transported from
large to small depths by some form of collective mean electromagnetic
action (Poynting flux). One should always keep in mind, however, that
this is by no means the only way in which energy can be removed
non-radiatively from the optically thick disc (for an alternative, see
e.g. Tagger \& Pellat 1999). 

We can now estimate the vertical Poynting flux, $F_{\rm P}$, 
in the simplest way, assuming that
$F_{\rm P}\simeq v_{\rm D} P_{\rm mag}$, where $v_{\rm D}$ is the
upward drift velocity of a magnetic flux tube within the disc. In
Merloni (2003) it was argued that $v_{\rm D}$ should in general be of
the order of the Alfv\'en speed $v_{\rm A}$. This translates into the following
expression for the fraction of power dissipated in the corona, uniquely
relating this quantity to the magnetic disc viscosity parameter
$\alpha_0$ \cite{mer03,hks06}:
\begin{equation}
\label{eq_f}
f \simeq \frac{v_{\rm A}}{c_{\rm s}}=
 \sqrt{2 \alpha_0 \beta^{\mu/2}},
\end{equation}
where $\beta = 1/(1+\xi)$, is the ratio of gas to total pressure, $\xi =
\pr/\pg$ is the ratio of the radiation to the gas pressure. Note that in this
approach $f$ is an implicit function of radius, through the radial dependence
of the pressures.

Recent progress in numerical studies of the disc-corona coupling
has been made by simulating a
gas-pressure dominated local patch of an accretion disc (with vertical gravity
included) in which heating by
dissipation of the MHD turbulence is balanced by radiative cooling (Hirose,
Krolik \& Stone 2006; see also Miller \& Stone 2000). 
In broad accordance with
eq. (\ref{eq_f}), it was found that the fraction of power released outside the
disc main body was less than about 10\% for a measured stress
parameter of $\alpha_0 \approx 0.02$. However, due to the increased magnetic
pressure support in the upper disc layers, most of the Poynting flux emerging
from the disc main body is dissipated {\it below} the photosphere, and
therefore cannot be directly associated with the observed hot, optically thin
X-ray emitting plasma. Obviously, global radiative 
simulations are needed to assess the role
of long-wavelength Parker instability modes, and the scaling with the 
radiation pressure predicted by eq. (\ref{eq_f}) for the generation of
 genuinely hot coronae from disc magentic fields.



\begin{figure}
\psfig{figure=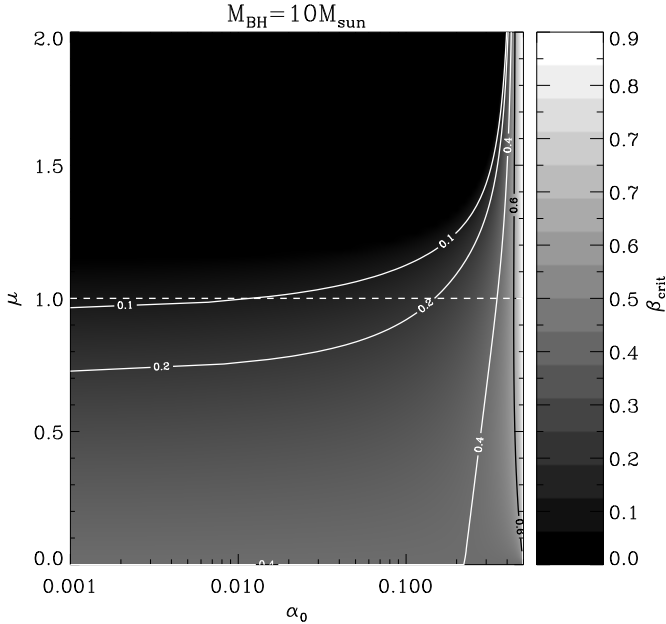,angle=0,width=0.5\textwidth}
\caption{The critical value of the gas to total pressure ratio, $\beta_{\rm
  crit}$, below which the disc is unstable as a function of the viscosity
  parameter $\alpha_0$ and index $\mu$. Calculations are for the case of no
  large scale magnetic field torques, and for a black hole mass $M_{\rm
  BH}=10M_{\odot}$. The solid lines mark the values $\beta_{\rm
  crit}=0.1,0.2,0.4,0.8$. The horizontal dashed line mark the $\mu=1$ case
  discussed in section~\ref{sec:visc}.}
\label{fig:beta}
\end{figure}

\section{Stability analysis}
\label{sec:stab}
Let us consider now the general case of the
viscosity prescription (\ref{eq:defmu}), with $0<\mu<2$. 
We can calculate analytically 
the value of the disc parameters at which the instability sets in
 by studying the stability properties of the
 stationary solution. In the one (vertical) zone limit, the equation for
hydrostatic equilibrium in the vertical direction is
\begin{equation} 
\label{eq:vert}
\ptot = {G M\Sigma H\over 2 R^3 }\; ,
\label{eq1} 
\end{equation}
while the angular momentum conservation equation reads:
\begin{equation}
\label{eq:ang}
P_{\rm gas}^{\mu/2} P_{\rm tot}^{(1-\mu/2)}= 
\frac{3 \Omega_{\rm K} \dot M J(R)}{8
    \pi \alpha_0 H}\,, 
\end{equation}
where the function $J(R)=(1-\sqrt{R_{\rm in}/R})$, with $R_{\rm in}=3
R_{\rm S}$, describes the Newtonian approximation of the no-torque at
the inner boundary condition for a disc around a Schwarzschild black
hole. Finally, the energy balance equation is given by
\begin{equation}
\label{eq:ene} 
\fr \simeq {c\pr\over \tau_T} = {3\over 2} \sigma_{r\phi}\, \Omega H=
\frac{3\Omega_{\rm K}^2 \dot M J(R) (1-f)}{8\pi}\,,
\label{eq2} 
\end{equation}
where $\fr$ is the vertical radiation flux and we have 
$f\propto \beta^{\mu/4}$, from eq.~(\ref{eq_f}).

By differentiating logarithmically the above expressions, together with
the equation of state~(\ref{eq:state}), with respect to $\dot M$, we
obtain four equations in terms of the logarithmic derivatives of
$P_{\rm tot}$, $\rho$, $T$ and $H$, that can be solved for $d \log
\rho / d \log \dot M$ and $d \log H/ d \log \dot M$ as functions of
$\beta$ and $f$. The lower turning point of the $\dot M(\Sigma)$ curve,
which indicates the local instability condition is then found by
setting
\begin{equation}
\frac{d\log \Sigma}{d\log \dot M}=\frac{d \log \rho}{d \log \dot
  M}+\frac{d \log H}{d \log \dot M}=0\,,
\end{equation}
which in turn gives:
\begin{equation}
\label{eq:instab}
\beta_{\rm crit}=\left(\frac{P_{\rm gas}}{P_{\rm tot}}\right)_{\rm crit}
=\frac{7\mu(2-3f)-16(1-f)}{7\mu(2-3f)-40(1-f)}\,.
\end{equation}
The instability sets in at a transition radius which can be found by solving
the following algebraic equation:
\begin{eqnarray}
\label{eq:rtr}
\frac{R_{\rm tr}}{J(R_{\rm tr})^{16/21}}&\simeq&350 R_{\rm S} 
\left(\frac{\beta_{\rm crit}}{1-\beta_{\rm crit}}\right)^{20/21}
\times \nonumber \\ && (\alpha_0
m)^{2/21} \dot m^{16/21} (1-f)^{6/7}
\end{eqnarray} 
where we have defined $m\equiv M_{\rm BH}/M_{\odot}$.

\begin{figure}
\psfig{figure=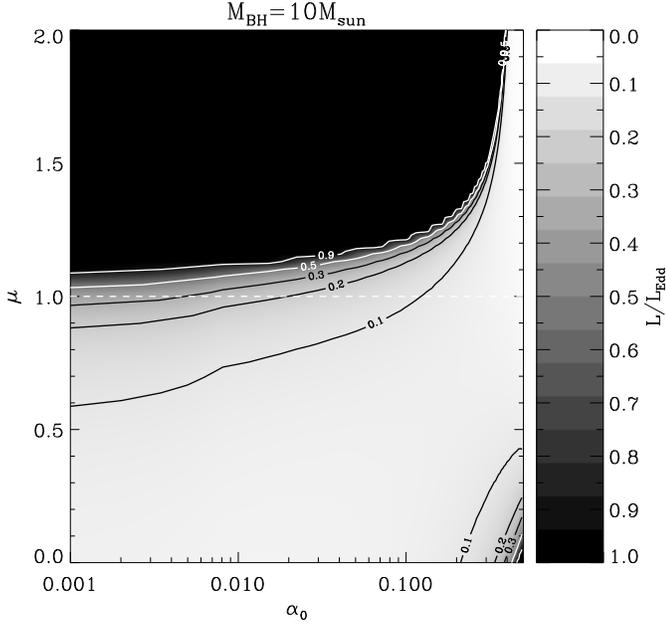,angle=0,width=0.5\textwidth}
\caption{The critical accretion rate in units of Eddington, $\dot
  m_{\rm crit}$, calculated
  for a radiative efficiency of $\epsilon_0=1/12$,
   above which a disc is unstable
is plotted as contours as a function of the
  viscosity parameter $\alpha_0$ and index $\mu$. Calculations are for a black
  hole mass $M_{\rm BH}=10M_{\odot}$. The solid
  lines mark the values $\dot m_{\rm crit}=0.15,0.2,0.3,0.5,0.9$. The
  horizontal dashed line mark the $\mu=1$ case discussed in section~\ref{sec:visc}.}
\label{fig:mdot}
\end{figure}

In the limit of $f=0$, we obtain the known result (see Szuszkiewicz 1990)
that, for the instability condition to be satisfied, the gas pressure needs to
be smaller than 0.4 times the total pressure if $\mu=0$, and just
$1/13$ times if $\mu=1$. 
Thus, the larger is $\mu$, the more difficult is to excite
the instability.  By using eq. (\ref{eq_f}) to relate $f$ and
$\beta$, 
we can then explore the whole space spanned by the parameters 
$\alpha_0$ and $\mu$.  In
Figure~\ref{fig:beta}, where we plot as contours the surface of critical gas
to total pressure ratio $\beta_{\rm crit}$, while in Figure~\ref{fig:mdot} we
plot the corresponding critical values of the accretion rate, defined
as $\dot m \equiv \epsilon \dot M c^2/L_{\rm Edd}$, where $\epsilon$
is the radiative efficiency\footnote{In the following, for the sake of
consistency with our Newtonian inner boundary
condition, we will assume $\epsilon=\epsilon_0\equiv 1/12$. 
However, it should be
kept in mind that for a black hole of arbitrary spin, the true
critical value of $L/L_{\rm Edd}$ should be rescaled by a factor
$\epsilon/\epsilon_0$ with respect to what shown in Fig.~\ref{fig:mdot}.}.

First of all, as it is well known, there exists a region of parameter
space for which the disc is always stable. Such a region correspond to
the upper left corners of figures~\ref{fig:beta} and~\ref{fig:mdot},
i.e. for large values of $\mu$, implying stresses coupled more with
the gas, rather than the total pressure, 
and small magnetic viscosity parameters
$\alpha_0$. 

On he other hand, if viscous stresses are proportional
to total pressure ($\mu=0$), the presence of a corona as a sink of
energy has a stabilizing effect on the disc \cite{sz94}, and the
critical accretion rate scales as $\dot m_{\rm crit}\propto
(1-f)^{-9/8}$. However, if $\alpha_0$ is large enough the disc can be
unstable even for large values of $\mu$ (upper right corner
of figures~\ref{fig:beta} and~\ref{fig:mdot}).

\subsection{Observational constraints}
It is an observational
fact that only the microquasar GRS 1915+105, the most
luminous transient black hole binary in our galaxy, displays strong
variability on timescales compatible with the limit cycle behaviour expected
from thermal instabilities (see Belloni et al. 1997;
Nayakshin et al. 2000; Janiuk et al. 2000,2002). Many other systems are known
to be stable up to luminosities which are a few tens of percent the Eddington
luminosity. This is clearly impossible for standard viscosity prescriptions
($\mu=0$), unless the viscosity parameter $\alpha_0$ (and thus the fraction of
power released into the corona) is high enough. But, as figures~\ref{fig:beta}
and~\ref{fig:mdot} show, there is only a very limited region of the
parameter space for which this is possible, and within such a region,
the fraction of power released outside the optically thick disc is
always large, in contradiction with the observed spectral properties of
high/soft state black hole binaries.

On the other hand, if $\mu\approx 1$, as argued, for example in
Merloni \& Fabian (2002) or Merloni (2003), 
we can have accretion discs which are
stable at more than half of the Eddington luminosity, and this holds
over a very large range in possible viscosity parameters
$\alpha_0$. This can indeed explain the observed lack of unstable,
disc-dominated galactic black hole binaries up to luminosities of
about a half of the Eddington one \cite{gd04}, while at the same time
allowing for {\it some} instability, taking in place only in those
systems which are constantly accreting very close to (or above) 
the Eddington rate, as supposedly is GRS 1915+105 \cite{dwg04}.

We can extend this result to the case of supermassive black
holes (SMBH). It is well known that for any value of $\mu$ and
$\alpha_0$ the critical accretion rate scales with mass as $\dot
m_{\rm cr} \propto M^{-1/8}$. Then for a $10^9$ solar masses black
hole the limit cycle instability should set in at a luminosity, in
units of Eddington, which is just one tenth of that of a binary black
hole, both viscosity law and black hole spin being the same. If
indeed most of the bright, high redshift Quasars that dominates the growth
history of SMBH have $L/L_{\rm Edd}>0.1$, as recently proposed
\cite{ves04,md04}, then they should be undergoing limit cycle
instability. 

\section{Time dependent evolution}
\label{sec:surf}
In order to test the above conclusions based on the stability analysis
of {\it stationary} solutions, we have carried out a set of time dependent
simulations for different values of the parameters $\alpha_0$ and
$\dot m$. From the equation of
conservation of mass and angular momentum, we can
write (Pringle, 1981; Livio \& Pringle 1992)
\begin{equation}
\frac{\partial \Sigma}{\partial t} = \frac{3}{R}
\frac{\partial}{\partial R} \left[ R^{1/2} \frac{\partial}{\partial R}
(\nu \Sigma R^{1/2}) \right],
\label{eqsigma}
\end{equation}

The time-dependent energy equation needed to close the set must take into
account large radial gradients of temperature, and therefore includes
a number of additional terms with respect to the standard heating and
cooling terms of the stationary solution. 
The form of the energy equation that we
use follows the formalism of Nayakshin et al. (2000):
\begin{eqnarray}
\ptot H \;{4-3\beta\over \Gamma_3 -1}\;\Bigg[ \left( {\partial \ln T
\over \partial t} \, + \,v_R {\partial \ln T
\over \partial R}\right) \nonumber \\
- \left(\Gamma_3-1\right) \,
\left({\partial \ln \Sigma\over \partial t} \, +\, v_R
{\partial \ln \Sigma\over \partial R} - {\partial H\over
\partial t}\right) \Bigg] \nonumber \\
= F^{+} - F^{-} - {2\over R}\, {\partial(R F_R H)\over
\partial R}  + J  ,
\label{eq5}
\end{eqnarray}
where $\gamma$ is the ratio of specific heats ($\gamma = 5/3$)
and $\Gamma_3$ is given in Abramowicz et al. (1995). Here, the radial
velocity $v_R$ induced by viscous stresses is given by (Eq. 5.7 of
Frank et al. 1992):
\begin{equation}
v_R = -{3\over \Sigma R^{1/2}}  \,{\partial\over \partial R}
\left[\nu\Sigma R^{1/2}\right].
\label{vr}
\end{equation}

The terms on the left hand side of Equation (\ref{eq5}) represent the
full time derivative (e.g., $\partial/\partial t + v_R
\partial/\partial R$) of the gas entropy, while the terms on the right
are the viscous heating, the energy flux in the vertical direction,
the diffusion of radiation in the radial direction, and the viscous
diffusion of thermal energy.  Following Cannizzo (1993; and references
therein), we take $J = 2 c_p
\nu (\Sigma/R) [\partial(R \partial T/\partial R)/ \partial R]$ to be
the radial energy flux carried by viscous thermal diffusion, where
$c_p$ is the specific heat at constant pressure.  $F^{+}$ is the
accretion disk heating rate per unit area, and is given by 
\begin{equation}
F^{+} = (9/4) \nu \Sigma \Omega_K^2 (1-f)\,.
\end{equation}
The radiation flux in the radial direction is
\begin{equation}
F_R = - 2 {c\pr\over \tau_T}\, H {\partial \ln T\over \partial R}\,,
\end{equation}
where $\tau_T$ is the optical depth of the disk, $\tau_T\equiv \kappa
\Sigma/2$, and $\kappa$ is the radiative opacity (assumed here to be dominated
by electron scattering opacity). Finally, the radiative cooling rate
in the vertical direction is given by
\begin{equation}
F^{-} = c\pr/\tau_T\,.
\end{equation}

\begin{figure}
\psfig{figure=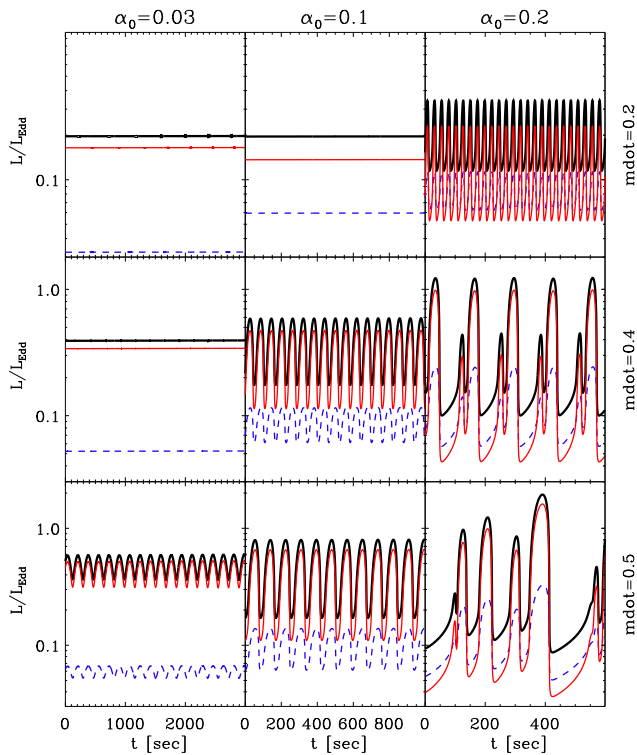,angle=0,width=0.50\textwidth}
\caption{Set of lightcurves for magnetized discs with the
   viscosity law corresponding to $\mu=1$, calculated for different
  accretion rates
  ($\dot m=0.2,0.4,0.5$) and viscosity parameters
  ($\alpha_0=0.03,0.1,0.2$). Each panel shows one lightcurve over a
  timescale approximately equal to one tenth of the viscous time at the outer
  boundary of our discs ($r=200$). Thick black lines represent the
  total (disc plus corona) emission, thin red lines the optically
  thick disc emission, and thin blue dashed ones the coronal (X-ray)
  emission. The black hole mass is fixed to 10$M_{\odot}$.}
\label{fig:nb}
\end{figure}

The results of our time-dependent simulations for the case $\mu=1$ and
 $M_{\rm BH}=14 M_{\odot}$ (as appropriate for GRS 1915+105) 
 are shown in figure~\ref{fig:nb}
 as a set of lightcurves plotted over a time typically of the order of the
 viscous time at the outer domain boundary.  Shown separately are the total
 disc (thin red solid lines) and coronal (blue dashed lines) 
emissions, together with
 their sum (thick black solid lines), 
 representing the total dissipated energy in the
 disc--corona system. These results broadly confirm the analysis presented in
 section~\ref{sec:stab}, in that we observe stable discs at $\dot m>0.4$ and
 $0.2$ for $\alpha_0=0.03$ and 0.1, respectively (see figure~\ref{fig:mdot}).

 We find in general a relationship between the amplitude of the instability
 and its duty-cycle (i.e. the ratio of the ``burst duration'' to the
 period of the oscillation), 
 whereas small duty cycles are associated with large
 amplitude variability. The amplitude itself is much smaller
 than in standard $\alpha$-discs, and grows with the ratio $\dot m/
 \dot m_{\rm crit}$. For accretion rates just above the critical
 values, the large cycles imply a ``refilling'' time of the inner disc
 faster than the viscous time at the transition radius. 
 In fact, we found that the
 viscosity prescription, $\mu \approx 1$, reproduces many
 features of the phenomenological one introduced by Nayakshin et
 al. (2000) in order to explain the time variability of GRS 1915+105
 (a task which is beyond the aim of this work).

Another general property of the simulated lightcurves is that, 
thanks to the nature of the
magnetic viscosity law adopted, the fraction $f$ of power dissipated
outside the optically thick disc is higher when
gas pressure dominates, i.e. when the inner disc is denser and cooler,
thus making coronal X-ray emission more
prominent during the low luminosity parts of the instability cycles.

\begin{figure}
\psfig{figure=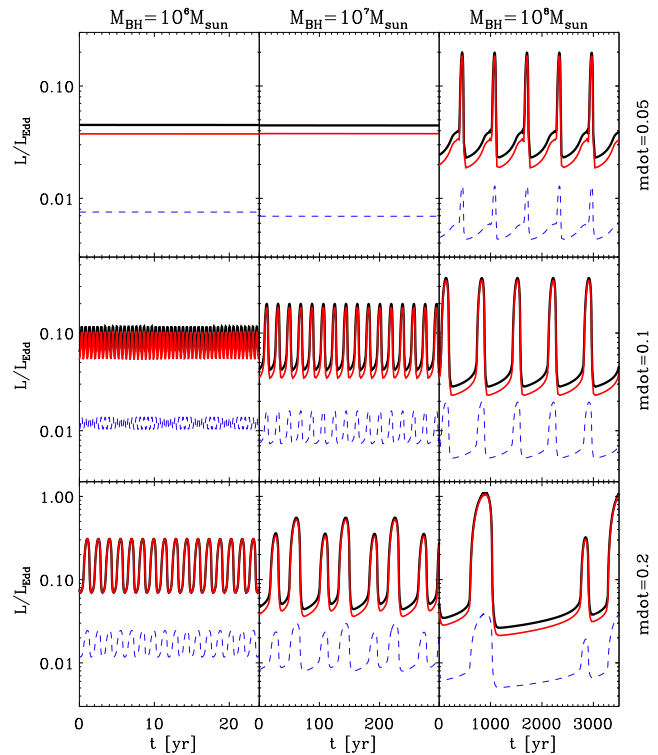,angle=0,width=0.50\textwidth}
\caption{Set of lightcurves calculated with $\mu=1$, $\alpha_0=0.03$,
  for different accretion rates
  ($\dot m=0.05,0.1,0.2$) and black hole masses
  ($M_{\rm BH}=10^6,10^7,10^8 M_{\odot}$). 
  Each panel shows one lightcurve over a
  timescale approximately equal to half of the viscous time at the outer
  boundary of our discs ($r=200$). Linestyles are as in Fig.~\ref{fig:nb}.}
\label{fig:nb_agn}
\end{figure}

Figure~\ref{fig:nb_agn} shows a set of simulated lightcurves
with $\alpha_0=0.03$ and $\mu=1$ for supermassive black holes of
different masses. As we noted in section~\ref{sec:stab}, for larger
black hole masses the instability sets in at lower $L/L_{\rm Edd}$. As
the duty cycle also gets smaller as the instability amplitude grows,
then we must conclude that the higher the external accretion rate,
$\dot M \propto M \dot m$, the more time will be spent by a source in
the dense, cold, low-luminosity portion of the instability cycles, 
close to its critical luminosity.

\section{Discussion}
\label{sec:discu}
The stability analysis we have presented here can be used to explain
the observed properties of accreting stellar mass 
black holes in different states.
First of all, a major constraint can be derived from the absence of any
limit-cycle instability in disc-dominated spectra. 
Those can be effectively selected in
two complementary ways: by detail modeling, choosing those spectra
in which the power-law component contribute to less than a fixed
amount (say, 10\%); and by studying the luminosity-temperature
relation of the disc component, selecting those spectra for which $L_{\rm
  disc} \propto T^4$ as expected from an optically thick disc extending
down to the innermost stable orbit (Kubota \& Makishima 2004). 
Modulo the distance uncertainty,
these studies unambiguously demonstrated that accretion discs can be
viscously stable up to at least $L_{\rm disc}/L_{\rm Edd} \ga 0.3$
\cite{gd04}.

From Figure~\ref{fig:mdot}, then, we conclude that only two disjoint 
 regions of parameter space are consistent with the observed
 phenomenology of black hole binaries. The first corresponds to 
$\alpha_0 \ga 0.3$ and, within our theoretical framework, it corresponds
 to accretion discs sandwiched by powerful, stabilizing coronae as
 postulated by Svensson \& Zdziarski (1994). The second acceptable
 region of the parameter space is roughly delimited by $\mu \ga 0.8$ 
and $\alpha_0 \la 0.15$ and would indicate that the stability
 properties of high/soft state black holes are dictated by a modified
 viscosity law.

The former constraint, however, implying 
 a large fraction of power transported vertically by
magnetic fields, seems inconsistent with the spectral properties
of the observed black holes in the high/soft state (but see however
Hirose et al. 2006 or Blaes et al. 2006). Moreover, for the only
source that exhibits limit-cycle instabilities, the microquasar GRS
1915+105, one can associate the observed variability time-scales with the
disc viscous time (Belloni et al. 1997; Nayakshin et
al. 2000; Janiuk et al. 2000). These models consistently suggest that
$\alpha_0 \sim $ a few percent. If this is indeed a general properties
of all accretion discs, a coherent picture emerges for black hole binaries in
the high/soft state in which MRI-driven turbulence generates stresses
which scale with the disc pressure as the geometric mean of gas and
 total pressure (Merloni 2003), i.e. $\mu
\approx 1$, with a constant of proportionality $\alpha_0$ of about a few
percent. This in turn implies quite low values of $f$ and consequent
small contribution from the power-law component to the observed X-ray
spectra.

\subsection{The low/hard state}
The scenario we have outlined above applies only to the high/soft state of
black hole binaries\footnote{For a general discussion of black hole binaries low/hard state, the reader is
referred to Markoff (2005) and Narayan (2005).}. It
seems unlikely that the soft-to-hard transition can be solely due to a
decrease of the external accretion rate and a consequent increase of the gas
pressure-dominated part of the disc, accompanied by an increase of the
strength of the corona (Merloni \& Fabian 2002). 
In fact, low values of $\alpha_0$ in the
high/soft state also imply low values of $f \approx \sqrt{2 \alpha_0}$ in the
low state, if the disc physics is not dramatically altered by the
$\dot m$ variation. 
Thus, there must be an additional physical mechanism, not included
in the simple treatment of the coupled magnetized disc+corona system presented
here, which is responsible for the observed state change. Obvious candidates
are: a radial transition to an inner optically thin, radiatively inefficient
flow, due to some kind of disc evaporation (Meyer \& Meyer-Hofmeister 1994;
Dullemond \& Turolla 1998; Spruit \& Deufel 2002), or a global rearrangement
of a large scale poloidal magnetic field, so that the accretion energy can be
dissipated almost entirely into the bulk flow of a relativistic jet (Livio,
Pringle \& King 2003). 

\subsection{Variability in AGN}
Since the radiation-pressure dominated regime sets in at smaller dimensionless
accretion rates for higher black hole masses, large scale variability due to
radiation-pressure driven viscous instabilities must be more widespread in AGN
than in stellar mass accreting black holes. However, observing such
variability in practice is only feasible for lower mass super-massive black
holes. Indeed, the viscous timescale, $t_{\rm visc}=(1/\Omega)(1/\alpha\beta^{\mu/2})(R/H)^2$, calculated at the transition radius (see
eq.~\ref{eq:rtr}) is approximately given by  
\begin{equation}
\label{eq:t_visc}
t_{\rm visc}(r_{\rm tr}) \approx 40 \alpha_0^{-2/3} m^{4/3} \dot
m^{2/3} \beta_{\rm crit}^{10/3-\mu/2} (1-\beta_{\rm crit})^{10/3}
\;\;{\rm s.}
\end{equation}  
This ranges from a few tens to a few hundreds of years if
$m$=$10^7$, consistent with estimates of switch on/off times for some
``changing look''  AGN (Guainazzi et al. 2005), to 
$10^{4}$--$10^5$ years for a $10^9$ solar masses black hole. This, 
interestingly, coincides with the
proposed intermittency time of radio galaxies, based on radio sources number
counts and on the properties of radio galaxies size distribution (Reynolds \&
Begelman 1997).

Another interesting aspect of the predicted variability for AGN
 concerns their feedback on galaxy formation. 
Indeed, in our model, black holes
accreting at a time-averaged sub-Eddington rate may spend a fraction of their
time accreting at above-Eddington accretion rate. If AGN feedback (for
 example in the form of powerful relativistic jets) is
significant when $L \simgt \ledd$, as suggested by the phenomenology
 of microquasars (see Fender, Belloni \& Gallo 2004), 
then such a source would produce feedback
while on the "hot" branch of the S-curve, whereas a completely stable source
at the same time-averaged accretion rate would not.

\section{Conclusions}

\label{sec:conc}

In this paper we discussed the implications of the
observed long term stability of the vast majority of black hole
X-ray binaries in the so-called high/soft state (or thermal dominant
state). This fact, together with the properties of the lightcurves 
of GRS 1915+105, the most luminous
galactic black hole and the only such system to
display limit-cycle-type of instabilities, can be used to put constraints on
the nature of the viscosity law in accretion discs.

We performed a stability analysis for accretion flows with
a flexible viscosity prescription under the hypothesis that the
Magneto-Rotational-Instability responsible for generating the
turbulent stresses inside the discs is also the source for a
magnetically dominated corona. By varying the scaling index of the
viscosity law, $\mu$, and its overall normalization, $\alpha_0$, we
have identified those regions of the parameter space for which
limit-cycle instability can develop as a function of the accretion
rate. From this analysis a coherent picture emerges for black hole binaries in
the high/soft state in which MRI-driven turbulence generates stresses
which scale with the disc pressure as the geometric mean of gas and
total pressure, i.e. $\mu
\approx 1$, with a constant of proportionality $\alpha_0$ of about a few
percent. This in turn implies quite low values of $f$ and consequent
small contribution from the power-law component to the observed X-ray
spectra. Provided that the viscosity law itself does not change
dramatically at lower accretion rates, a consequence of this
 result is that the transition to the low/hard state must be caused by
 additional physics, such as the evaporation of, or a global
 large scale magnetic field re-arrangement in the inner portions of
 the disc.

The scaling with black hole mass of the accretion equations is such
that limit cycle instabilities should play a role in all bright AGN
and Quasars. The typical timescale for these oscillations, however,
grows with black hole mass faster than linearly.
Only smaller mass black holes (less than a few times $10^7$ solar
masses) may have limit cycle instability timescales
of just a few years. Even considering the uncertainties in the absolute
values of the predicted critical luminosity that come from
uncertainties in the black hole mass and spin, it 
would be interesting to search for any evidence of the predicted
behaviour by looking at distribution of observed disk luminosities in
large samples of AGN with available estimates of the central black hole
mass.


\bsp

\label{lastpage}


\begin{thebibliography}{}



\bibitem[Abramowicz et al. 1988]{abr88}
Abramowicz, M. A., Czerny, B., Lasota, J. P., Szuszkiewicz, E., 1988, ApJ, 332, 646.


\bibitem[Abramowicz et al. 1995]{abra95}
Abramowicz M. A., Chen X. \& Taam R., 1995, ApJ, 452, 379

\bibitem[Balbus \& Hawley 1998]{bh98}
Balbus, S. A. \& Hawley, J. F., 1998, Rev. Mod. Phys. 70, 1.

\bibitem[Belloni et al. 1997]{bel97}
Belloni T., M\'endez M., King A. R., van der Klis M., van Paradjis J.,
1997, ApJL, 488, L109


\bibitem[Blackman \& Field 2000]{bf00}
Blackman E. C. \& Field G. B., 2000, MNRAS, 318, 724


\bibitem[Blaes et al. 2006]{bla06}
Blaes O. M., Davis S. W., Hirose S., Krolik J. \& Stone J. M., 2006,
ApJ, 645, 1402


\bibitem[Blaes \& Socrates 2001]{bs01}
Blaes, O. \& Socrates, A., 2001, ApJ, 553, 987.


\bibitem[Burm 1985]{bur85}
Burm, H., 1985, A\&A, 143, 389.



\bibitem[Cannizzo 1993]{can93}
Cannizzo J. K., 1993, ApJ, 429, 318


\bibitem[Done, Wardzi\'nski \& Gierli\'nski 2004]{dwg04}
Done C., Wardzi\'nski G.  \& Gierli\'nski M., 2004, MNRAS, 349, 393



\bibitem[Dullemond \& Turolla 1998]{dt98}
Dullemond C. P. \& Turolla R., 1998, ApJ, 503, 361



\bibitem[Fender \& Belloni 2004]{fb04}
Fender R. P., Belloni T., 2004, ARA\&A 42, 317



\bibitem[Fender, Belloni \& Gallo 2004]{fbg04}
Fender R. P., Belloni T. \& Gallo E., 2004, MNRAS, 355, 1105



\bibitem[Frank, King and Raine 1992]{fkr92}
Frank J., King A., Raine D., 1992, Accretion Power in Astrophysics
(Cambridge: Cambridge Univ. Press)



\bibitem[Galeev, Rosner \& Vaiana 1979]{grv79}
Galeev, A. A., Rosner, R. \& Vaiana, G. S., 1979, ApJ, 229, 318


\bibitem[Gierli\'nski \& Done 2004]{gd04}
Gierli\'nski, M. \& Done C., 2004, MNRAS, 347, 885


\bibitem[Guianazzi et al. 2005]{gua05}
Guainazzi M., Fabian A. C., Iwasawa K., Matt G., Fiore F., 2005, MNRAS, 356, 295



\bibitem[Haardt \& Maraschi 1991]{hm91}
Haardt, F. \& Maraschi, L., 1991, ApJL, 380, L51.



\bibitem[Hirose, Krolik \& Stone 2006]{hks06}
Hirose S., Krolik J. H. \& Stone J M., 2006, ApJ, 640, 901



\bibitem[Honma, Matsumoto \& Kato 1991]{hmk91}
Honma F., Matsumoto R. \& Kato S., 1991, PASJ, 43, 147



\bibitem[Hopkins et al. 2006]{hop06}
Hopkins P. F., Narayan R., Hernquist L., 2006, ApJ, 643, 641



\bibitem[Janiuk, Czerny, Siemiginowska 2000]{jcs00}
Janiuk A., Czerny B. \& Siemiginowska A., 2000, ApJL, 542, L33



\bibitem[Janiuk, Czerny, Siemiginowska 2002]{jcs02}
Janiuk A., Czerny B. \& Siemiginowska A., 2002, ApJ, 576, 908



\bibitem[Klein-Wolt et al. 2002]{kw02}
Klein-Wolt M., Fender R. P., Pooley G. G., Belloni T., Migliari S.,
Morgan E. H., van der Klis M., 2002, MNRAS, 331, 745




\bibitem[Kubota \& Makishima 2004]{km04}
Kubota A. \& Makishima K., 2004, ApJ, 601, 428





\bibitem[Kuncic \& Bicknell 2004]{kb04}
Kuncic Z. \& Bicknell G. V., 2004, ApJ, 616, 669


\bibitem[Liang \& Price 1977]{lp77}
Liang E. P. T. \& Price R. H., 1977, ApJ, 218, 247 



\bibitem[Lightman \& Eardley 1974]{le74}
Lightman A. P. \& Eardley D. M., 1974, ApJ, 187, L1


\bibitem[Livio \& Pringle 1992]{lp92}
Livio M. \& Pringle J. E., 1992, MNRAS, 259, L23



\bibitem[Livio, Pringle \& King 2003]{lpk03}
Livio M., Pringle J. E. \& King A. R., 2003, ApJ, 593, 184


\bibitem[Markoff 2005]{mar05}
Markoff S., 2005, Ap\&SS, 300, 189



\bibitem[McClintock \& Remillard 2006]{mr06}
McClintock J. E. \& Remillard R. A., 2006, to appear in ``Compact
Stellar X-ray Sources'', eds. W.H.G. Lewin and M. van der Klis. astro-ph/0306213



\bibitem[McLure \& Dunlop 2004]{md04}
McLure M. J. \& Dunlop J. S., 2004, MNRAS, 352, 1390


\bibitem[Merloni 2003]{mer03}
Merloni A., 2003, MNRAS, 341, 1051



\bibitem[Merloni 2004]{mer04}
Merloni A., 2004, MNRAS, 353, 1035


\bibitem[Merloni \& Fabian 2002]{mf02}
Merloni, A. \& Fabian, A. C., 2002, MNRAS, 332, 165



\bibitem[Meyer \& Meyer-Hofmeister 1994]{mm94}
Meyer F. \& Meyer-Hofmeister E., 1994, A\&A, 361, 175



\bibitem[Miller \& Stone 2000]{ms00}
Miller K. A. \& Stone J. M., 2000, ApJ, 534, 398


\bibitem[Narayan 2005]{nar05}
Narayan R., 2005, Ap\&SS, 300, 177



\bibitem[Nayakshin, Rappaport \& Melia 2000]{nay00}
Nayakshin S., Rappaport S., Melia F., 2000, ApJ, 535, 798



\bibitem[Pringle 1981]{pri81}
Pringle J. E, 1981, ARA\&A, 19, 137



\bibitem[Pringle, Rees \& Pacholczyk 1973]{prp73}
Pringle J. E., Rees M. J. \& Pacholczyk A. G., 1973, A\&A, 29, 179





\bibitem[Reynolds \& Begelman 1997]{rb97}
Reynolds C. S. \& Begelman M. C., 1997, ApJL, 487, L135


\bibitem[Sano et al. 2004]{sano04}
Sano T., Inutsuka S., Turner N. J. and Stone J. M., 2004, ApJ, 605, 321


\bibitem[\protect\citename{Shakura \& Sunyaev }1973]{ss73}
Shakura, N. I. \& Sunyaev, R.A., 1973, A\&A, 24, 337.




\bibitem[Spruit \& Deufel 2002]{sd02}
Spruit H. C.\& Deufel B., 2002, A\&A, 387, 914



\bibitem[Stella \& Rosner 1984]{sr84}
Stella, L. \& Rosner, R.,  1984, ApJ, 277, 312.



\bibitem[\protect\citename{Svensson \& Zdziarski }1994]{sz94}
Svensson, R. \& Zdziarski, A. A., 1994, ApJ, 436, 599.



\bibitem[\protect\citename{Szuszkiewicz }1990]{szu90}
Szuszkiewicz, E., 1990, MNRAS, 244, 377



\bibitem[\protect\citename{Szuszkiewicz \& Miller }1997]{sm97}
Szuszkiewicz E. \& Miller J., 1997, MNRAS, 287, 165



\bibitem[\protect\citename{Taam \& Lin }1984]{tl84}
Taam, R. E. \& Lin, D. N. C., 1984, ApJ, 287, 761

\bibitem[\protect\citename{Tagger \& Pellat }1999]{tp99}
Tagger M. \& Pellat R., 1999, A\&A, 349, 1003



\bibitem[Turner, Stone \& Sano 2002]{tss02}
Turner, N. J., Stone, J. M. \& Sano, T.,  2002, ApJ, 566, 148.



\bibitem[Vestergaard 2004]{ves04}
Vestergaard M., 2004, ApJ, 601, 676



\bibitem[Watarai \& Mineshige 2003]{wm03}
Watarai K.-Y. \& Mineshige S., 2003, ApJ, 596, 421



\bibitem[Yu \& Tremaine 2002]{yt02}
Yu Q. \& Tremaine S., 2002, MNRAS, 335, 965 (YT03)



\end{thebibliography}
\end{document}